\documentclass[numberedappendix]{emulateapj}
\usepackage{stmaryrd}
\usepackage[USenglish]{babel}
\usepackage{amsmath,amssymb,amsxtra,amsfonts}
\usepackage{graphicx}
\usepackage{epstopdf}
\usepackage{hyperref}
\usepackage{natbib}
\usepackage{bm}
\usepackage{color}
\usepackage{enumerate}

\newcommand{\ba}{\begin{eqnarray}}
\newcommand{\ea}{\end{eqnarray}}
\newcommand{\be}{\begin{equation}}
\newcommand{\ee}{\end{equation}}
\newcommand{\bs}{\begin{split}}
\newcommand{\es}{\end{split}}
\newcommand{\bi}{\begin{itemize}}
\newcommand{\ei}{\end{itemize}}
\newcommand{\flcdm}{flat $\Lambda$CDM}
\newcommand{\fltb}{Lema\^{\i}tre-Tolman-Bondi}

\shortauthors{H. Wang and T. Zhang}
\shorttitle{Testing LTB with OHD}

\begin{document}
\title{Constraints on Lema\^{\i}tre-Tolman-Bondi models from Observational Hubble Parameter data}

\author{Hao Wang\altaffilmark{1}, Tong-Jie Zhang\altaffilmark{1, 2}}
\altaffiltext{1}{Department of Astronomy, Beijing Normal University, Beijing
100875, China}
\altaffiltext{2}{Center for High Energy Physics, Peking University, Beijing
100871, China}
\email{tjzhang@bnu.edu.cn}

\begin{abstract}
We use the observational Hubble parameter data (OHD), both the latest observational dataset (\cite{Stern2010}, referred to as SJVKS) and the simulated datasets, to constrain {\fltb} (LTB) void models. The necessity of the consistency check on OHD itself in the LTB cosmology is stressed. Three voids are chosen as test models and are constrained using the Union2 dataset of SN Ia as well as OHD. Despite their different parametrization, the results from our test models show some indicating similarities, e.g., the best-fit voids obtained from OHD are all considerably broader than those from SN Ia. Due to the small size of the SJVKS dataset, the constraints are not conclusive. The constraining power of the future OHD observations are therefore investigated, through a Figure of Merit (FoM) analysis based on the Monte Carlo simulated data. We found that, in the case that the future OHD become more consistent with SN Ia, the results from the test models are almost unanimous: 1) as many as $32$ OHD data points at the SJVKS-like uncertainty level are needed to give a higher FoM than the Union2 dataset of SN Ia; 2) precise observation helps reduce this required number; 3) increasing the survey depth does not always increase the FoM. On the other hand, if the future OHD and the Union2 dataset keep favor different voids, in a similar manner as they do at present, the $1\sigma$ confidence regions obtained from the two probes should finally separate. We test this conjecture and found that, the minimum observational requirement (the size of the dataset, the uncertainty level and the survey depth) for this inconsistency to emerge depends strongly on the void model.
\end{abstract}

\keywords{cosmology:miscellaneous; statistical --- methods: numerical}

\section{Introduction}
Since the direct supernovae measurement of the deceleration parameter \citep{Riess1998, Perlmutter1999}, the last decade has seen a solidification of the standard cosmological model, which has about $75\%$ dark energy driving the acceleration of a flat Friedmann-Robertson-Walker (FRW) universe. A very long array of observations, including cosmic microwave background (CMB) \citep{Komatsu2011}, baryon acoustic oscillations (BAO) \citep{Percival2010}, weak lensing \citep{Schrabback2010}, etc., are consistent with this model.

Yet a radically different scenario has attracted considerable attention recently as an alternative model choice to fit the observations. The idea is to suppose we are very near to the center of a large spherical void, which is itself embedded in an otherwise homogeneous background \citep{celerier2000}. The dynamics of this kind of models can be described in Lema\^{i}tre-Tolman-Bondi (LTB) metric, hence the term LTB models. Constraints from supernovae, CMB, BAO, and $H_0$ have shown that this model can not be ruled out (\cite{Garcia-Bellido2008, Biswas2010}, but see \cite{Moss2011} and \cite{Zibin2008a}). As the cross-check from the consistency of the results based on independent evidences is an effective, if not the only, way to pin down the correct cosmological model \citep{Peebles2003}, more observations are needed to better discriminate the standard FRW and the LTB scenarios.

Observational Hubble parameter data (OHD) has been measured through the aging of passively evolving galaxies \citep{Jimenez2002, Simon2005} and BAO effect \citep{Gaztanaga2009}. Recently, a third method has been proposed to obtain OHD from the gravitational wave standard sirens \citep{Nishizawa2011}. OHD is shown to be consistent with the standard cosmological model \citep{Lin2009, Stern2010}. Various other FRW models have also been tested against OHD (\cite{Samushia2006, Yi2007, Zhang2008, Zhai2010, Xu2010}, see \cite{Zhang2010a} for a recent review). The potential of future OHD observations in constraining FRW models is explored via simulation in \cite{Ma2011}. Only recently were this dataset used to test LTB models in \cite{February2010} (hereafter FLSC).

The current work deals with a similar purpose as did in FLSC, as we believe more discussions are needed, both on the subtleties when using OHD in LTB cosmology, and on its potential constraining power as the dataset enlarges in the future. The investigations on these two aspects, as well as the current constraints on various void models, constitute the main content of this paper.

We briefly review the LTB cosmology and introduce several test models in Section \ref{sec.ltb}. In Section \ref{sec.ohd} we present the physics and assumptions involved in obtaining OHD. This analysis then guides our choice in picking the proper observable in LTB cosmology correspondent to the OHD dataset, and signifies the necessity for the consistency check on OHD itself in the LTB cosmology. Section \ref{sec.current} illustrates how, in the LTB cosmology, the likelihood analysis should be done when OHD is used; the best-fit void parameters obtained in this section are used in Section \ref{sec.vslcdm} to compare between the LTB void models and the standard model. The potential constraining power of the future OHD observations is explored in Section \ref{sec.forecast}, through a Figure of Merit analysis; in the same section we also discuss whether the future OHD dataset can be used, when combined with the Union2 dataset, to exclude the void models. We conclude in Section \ref{sec.conclusion}.

\section{LTB Models}\label{sec.ltb}
The LTB solution to the Einstein field equations describes the dynamics of a spherically symmetric dust universe. We take a brief review of the basic formulas in LTB cosmology, then we introduce the test LTB models used in our analysis.
\subsection{Basics of LTB Cosmology}
The background LTB line-element is written as:
\be
\label{Eq.ltbmetric}ds^2=-dt^2+\frac{A'(r, t)^2}{1-k(r)}dr^2+A^2(r, t)d\Omega^2,
\ee
where $'$ denotes $\partial / \partial r$, and $k(r)$ is an arbitrary function of $r$, playing the role of spatial curvature. The stress-energy tensor of the mass source is given by:
\be
T_\mu^\nu=-\rho(r, t)\delta_0^\nu\delta_\mu^0.
\ee

Note that the FRW metric can be recovered by imposing $A(r, t)=a(t)r$ and $k(r)=kr^2$. The Einstein field equations read:
\begin{align}
\label{Eq.Fried1}H_{\bot}^2+2H_{\bot}H_{\parallel}+\frac{k(r)}{A^2}+\frac{k'(r)}{AA'} = 8\pi G\rho_m,\\
\label{Eq.Fried2}\dot A^2+2A\ddot A+k(r) = 0,
\end{align}
where $\dot{}$ denotes $\partial / \partial t$, and
\begin{align}
\label{Eq.Hperp}H_{\bot}\equiv\frac{\dot A^2}{A^2},\\
\label{Eq.Hparr}H_{\parallel}\equiv\frac{\dot A'}{A'},
\end{align}
are the expansion rates at the transverse and longitudinal direction, respectively. Integrating Eq.\ref{Eq.Fried2}, we get
\be
\label{Eq.Hper}H_{\bot}^2=\frac{M(r)}{A^3}-\frac{k(r)}{A^2},
\ee
where $M(r)$ is another arbitrary function. The evolution of $A(r,t)$ can be obtained by integrating Eq.\ref{Eq.Hper}; the standard parametric solutions are as follows \citep{Moss2011}:
\ba
\label{Eq.ps1}
\begin{split}
A(r, t)=&\frac{M(r)}{k(r)}(1-\text{cosh}{\eta}),\\
t-t_B(r)=&\frac{M(r)}{(-k(r))^{3/2}}(\text{sinh}\eta-\eta),
\end{split}
\ea
for $k<0$;
\ba
\label{Eq.ps2}
\begin{split}
A(r, t)=&\frac{M(r)}{k(r)}(1-\text{cos}\eta),\\
t-t_B(r)=&\frac{M(r)}{k(r)^{3/2}}(\eta-\text{sin}\eta),
\end{split}
\ea
for $k>0$; and
\be
\label{Eq.ps3}A(r, t)=\left(\frac{9M(r)}{2}\right)^{1/3}\left(t-t_B(r)\right)^{2/3},
\ee
for $k=0$. Here $t_{B}$ is another arbitrary function, referred to as the `bang time', due to the singularity behavior at $t= t_{B}$.

Substituting Eq.\ref{Eq.Hper} into Eq.\ref{Eq.Fried1} gives
\be
\label{Eq.Mprime}\frac{M'(r)}{A'A^2}=8\pi G\rho_m.
\ee

Evaluating Eq.\ref{Eq.Hper} at present time gives the interdependence between $M(r)$ and $k(r)$, which, with a notation similar to that in FRW cosmology, can be parameterized as:
\be
\begin{split}
\label{Eq.Mk}M(r)=&H_{\bot 0}^2(r)\Omega_m(r) A_0^3(r),\\
k(r)=&H_{\bot 0}^2(r)[\Omega_m(r)-1]A_0^2(r),
\end{split}
\ee
where a subscript $0$ denotes the current value ($t=t_0$) of a quantity. Substituting Eq.\ref{Eq.Mk} and $A(t=t_0, r)=A_0(r)$ to Eq.\ref{Eq.ps1} (or Eq.\ref{Eq.ps2}, Eq.\ref{Eq.ps3}, depending on the value of $k(0)$), one can get the age of the universe as a function of $r$, given by
\be
\label{Eq.Age}t_0-t_B(r)=\frac{1}{H_{\bot 0}(r)}\mathcal{F}(\Omega_m),
\ee
where
\be
\label{Eq.mathf}\mathcal{F}(x)=
\begin{cases} \frac{- \sqrt{x-1}+x\sin^{-1} \sqrt{\frac{x-1}{x}}} {\left(x-1\right)^{3/2}}   &  x>1 \\ 2/3 & x=1 \\
\frac{\sqrt{1-x} - x\sinh^{-1}\sqrt{\frac{1-x}{x}} } {(1-x)^{3/2}}   &  x<1
\end{cases}
\ee

In order to compare the theory with the observation, one needs to associate the coordinates with the redshifts. This can be done by noticing the redshift equation \citep{Alnes2006, Enqvist2008}
\be
\label{Eq.dzdt}\frac{d\ln (1+z)}{dt}=-\frac{\dot A'(r, t)}{A'(r, t)},
\ee
and the equation of $dt$ and $dr$ on the light cone
\be
\label{Eq.dtdr}\frac{dt}{dr}=-\frac{A'(r, t)}{\sqrt{1-k(r)}}.
\ee

As a Cauchy problem the differential equations can be solved numerically for any redshift, once the profile of $\Omega_m$ and $H_{\bot0}$ are set. We do the equation integration with a modified open-source software easyLTB \citep{Garcia-Bellido2008}, where $A_0(r)=r$ is chosen to fix the gauge. The calculation process for the cases when $\Omega_k\neq 0$ is as following,
\begin{enumerate}[1)]
\item Start from $t=t_0$, $r=0$;
\item\label{It.eta} Solve for $\eta$ from the second equation of Eq.\ref{Eq.ps1} or Eq.\ref{Eq.ps2};
\item Obtain the expression of $A$, $\dot{A}$ and $A'$ from the first equation of Eq.\ref{Eq.ps1} or Eq.\ref{Eq.ps2}. $\eta'$ and $\dot\eta$ in the result can be expressed as functions of $A$ using the second equation;
\item Solve $\dot{A}'$ from Eq.\ref{Eq.Fried1}, Eq.\ref{Eq.Hperp}, Eq.\ref{Eq.Hparr}, Eq.\ref{Eq.Mprime}, and Eq.\ref{Eq.Mk};
\item\label{It.integ} Integrate Eq.\ref{Eq.dzdt} and Eq.\ref{Eq.dtdr} over $z$ to get the new $t(z)$ and $r(z)$;
\item Iterate steps \ref{It.eta}-\ref{It.integ} till the desired $z$.
\end{enumerate}

The original easyLTB does not support the void models with $\Omega_k\leq0$. We modify the code to allow for negative $\Omega_k$. For $\Omega_k=0$, it is not valid anymore to use the parametric forms. However, as both Eq.\ref{Eq.ps1} and Eq.\ref{Eq.ps2} converge to Eq.\ref{Eq.ps3} when $|\Omega_k|\rightarrow 0$, in practice one can always approximate the correct answer with the parametric form by setting $\Omega_k$ to a small nonzero value: following the calculation process above, letting $\Omega_k=10^{-6}$ would cause a $10^{-3}$ error in $\eta$ and lead in turn to an $10^{-9}$ error in $A$ for each integration step; the error will accumulate during the integration but should not be larger than $1\%$ for any sensible number of integration steps (normally smaller than 100000).

One can tell the angular diameter distance directly from the metric form, Eq.\ref{Eq.ltbmetric},
\be
d_A(z)=A\left(r(z), t(z)\right),
\ee
hence the luminosity distance (the distance duality relation holds as long as one stays with the metric theory of gravity and does not consider any exotic physics \citep{Etherington1933, Bassett2004}),
\be
d_L(z)=(1+z)^2A\left(r(z), t(z)\right).
\ee

The LTB dynamics discussed above consider only matter, so they break down at high redshifts where radiation become important. This is one reason why a large family of LTB models are voids that are embedded in an FRW background universe \citep{Biswas2010, Zibin2008a, Moss2011}. Moreover, out of the consideration of meeting the prediction of the inflation, an Einstein-de Sitter universe is often chosen as this background \citep{Garcia-Bellido2008, February2010}. The LTB models so constructed are referred to as EdS voids. This type of voids are said to be ruled out as a whole according to \cite{Moss2011} (see also \cite{Zibin2008a}), as those voids which fit the CMB data will give too low a $H_0$ that will contradict the local observation. \cite{Clarkson2011} found, however, that there are ways out if one takes into account the radiation when joining the voids to the background FRW universe, for instance, by introducing more relativistic degrees of freedoms, or by considering varied baryon fraction and/or baryon-photon ratio.

Our aim in this work is mainly in exploring the constraining power of OHD data, compared with SN Ia data, both of which are low redshift observations. Therefore we do not have to take into account the role played by the radiation. Besides, whether the EdS voids are feasible or not, as test models they could in any case serve as a test bench of the constraining power of different observational probes. Thirdly, as mentioned above, the EdS background is consistent with inflation, and is therefore a theoretically conservative choice. For these reasons, the test models we use in this work are all EdS voids. Also, we assume there are no isocurvature modes, i.e., variation in the baryon fraction or in the baryon-photon ratio, correlated with the void, because these will enhance the inhomogeneity in the structure formation time which might make OHD itself invalid, not to mention using it to constrain models (see the discussions in Section \ref{sec.ohd}).

\subsection{Void Models}
The radial inhomogeneity of LTB solutions leaves an almost totally arbitrary form of the radial profiles of $\Omega_m$, $H_{\bot 0}$ and $t_B$. For $\Omega_m$, the much discussed void-like profiles are favored. This is because of the requirement that the LTB model should have an expansion rate decreasing with the distance to the center, which in turn resemble observationally an expansion rate increasing with time in FRW models.

Besides, the gradients in the bang time, $t_B$, corresponds to a currently non-vanishing decaying mode \citep{Silk1977, Zibin2008}. This would imply a very inhomogeneous early universe, hence violate inflation. More importantly for the current work, this will lead to great inhomogeneities in the galaxy formation time and make the OHD dataset invalid. So $t_B$ must be a constant, and we set it to be $0$. $H_{\bot 0}$ for the desired void models is then given by substituting $t(r)\equiv t_0$ into Eq.\ref{Eq.Age}:
\be
\label{Eq.modelh}H_{\bot0}(r)=H_0\mathcal{F}(\Omega_m),
\ee
where
\be
\label{Eq.H0_Age}H_0\equiv 1/t_0,
\ee
and $\mathcal{F}$ is given by Eq.\ref{Eq.mathf}.

\begin{figure}[!htb]
\begin{center}
\includegraphics[width=\columnwidth]{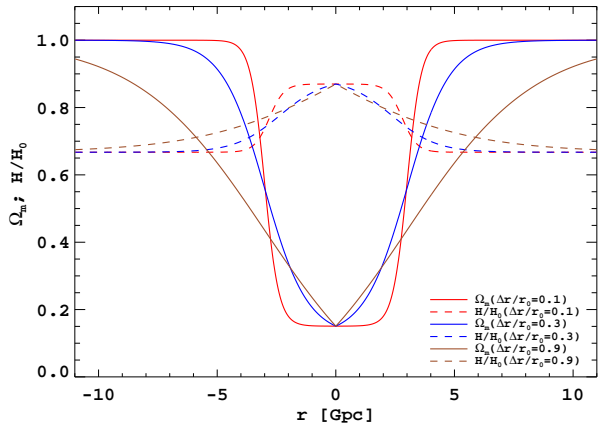}
\end{center}
\caption{$\Omega_m$ and $H_{\bot 0}$ as functions of $r$, with fixed $H_0$, $\Omega_0$, $r_0$, and different $dr$. Small $dr$ is seen to lead to profiles that are flat near the origin, and large $dr$ more pointed ones.
 \label{Fig.om_h}}
\end{figure}

With the relation between $H_{\bot0}$ and $\Omega_m$ determined by Eq.\ref{Eq.modelh}, we need only fix the specific profile of $\Omega_m$. The Constrained GBH (CGBH) model \citep{Garcia-Bellido2008} is chosen to be one of our test models,
\be
\label{Eq.modelom}\Omega_m(r)=1+(\Omega_{0}-1)\left[\frac{1-\tanh[(r-r_0)/2\Delta{r}]}{1+\tanh(r_0/2\Delta{r})}\right],
\ee
where $\Omega_{0}$ describes the density at the symmetric center, $r_0$ is the characteristic size of the void, and $\Delta r$ describes the steepness of the void near the edge. Figure \ref{Fig.om_h} illustrates profiles of $\Omega_m$ and $H_{\bot 0}$ in the CGBH model, with $H_0$, $r_0$, and $\Omega_0$ fixed at 74 kms$^{-1}$Mpc$^{-1}$, 0.05, 2Gpc, respectively. Different colors (red, blue, brown) stand for different values of $\Delta r/r_0$ (0.1, 0.3, 0.9, respectively). It can be seen that small $\Delta r/r_0$ corresponds to an $\Omega_m$ profile that is flat at the origin, i.e., $\Omega_m'|_{r=0}=0$, while large $\Delta r/r_0$ more pointed ones ($\Omega_m'|_{r=0}\neq0$). Because of this feature, and also the consideration that $\Delta r$ should not be greater than unity, we use $dr\equiv \Delta r/r_0$ as a free parameter instead of $\Delta r$. Note that despite the notation, $\Omega_m$ does not reflect the exact profile of the matter density. At a given cosmic time $t^*$, the density profile $\rho_m(r, t^*)$ is determined through Eq.\ref{Eq.Fried1}.

The value of $\Omega_m'$ at the origin turns out to be of great importance. Defining the curvature as $K\equiv k/r^2$, one could tell from Eq.\ref{Eq.Mk} and Eq.\ref{Eq.modelh} that the r-derivative of $K$, $K'$, is directly related to $\Omega_m'$. Indeed, it is straightforward to show that an LTB model with homogeneous bang time is smooth at the center, $K'|_{r=0}=0$, if and only if $\Omega_m'|_{r=0}=0$ and $\frac{d\mathcal{F}}{d\Omega_m}|_{r=0}$ is finite. This kind of voids are referred to here as the `smooth voids'. Particularly, CGBH void is a smooth void only when $dr\rightarrow 0$. It has been argued that the smooth voids are more natural. More importantly, the luminosity distance-redshift relation in the smooth voids differs qualitatively from that of the FRW models, which could be used to differentiate these two scenarios with the future supernovae data \citep{Clifton2008}. Keeping this result in mind, we choose the rest of our test models to be always smooth. From the discussion above, this is to look for models with $\Omega_m$ that is differentiable at the origin. The first such model we choose has the simplest Gaussian profile:
\be
\label{Eq.gaussom}\Omega_m(r)=1+(\Omega_{0}-1)\text{exp}\left(-\frac{r^2}{2r_0^2}\right),
\ee
with the parameter set as \{$\Omega_{0}$, $r_0$, $H_0$\}. The second smooth model is obtained through a cubic spline interpolation. The constraints we set, given three $r$ values $r_1=0<r_2<r_3$, for determining the $\Omega_m$ profile, are that
\begin{enumerate}[(a)]
 \item $\Omega_m(0)=\Omega_0$;
 \item $\Omega_m'|_{r=0}=0$, i.e., a smooth void;
 \item $\Omega_m(r_2)=\Omega_{r2}$;
 \item $\Omega_m(r_3)=1$;
 \item $\Omega_m'|_{r=r_3}=0$, which together with (d) makes this model an EdS void.
\end{enumerate}
The parameter set for the Spline model is then \{$\Omega_0$, $r_2$, $\Omega_{r2}$, $r_3$, $H_0$\}. In practice, to make sure $r_2$ is no larger than $r_3$ we use $r_2/r_3$ as a free parameter instead of $r_2$,

Throughout the paper the CGBH model is treated as our main model. Most of the calculation details will be described only for this model, although the main results for all three models will always be shown and discussed together.
 	
\section{Observational Hubble Parameter From Passively evolving Galaxies}\label{sec.ohd}
In this section, we firstly give a brief review on how OHD is obtained and utilized in the FRW scenario, then we will figure out the complexities that arise when moving to the LTB scenario. As mentioned in the introduction, there are currently two sources of OHD data points, those from the passively evolving galaxies and those from the large scale structure (BAO). The latter method depends on the detailed evolution of perturbations which is not well understood in the LTB cosmology, although progresses have been made \citep{Zibin2008, Clarkson2009}. Therefore, the OHD discussed in this work refer exclusively to the former.
\subsection{FRW scenario}
The Hubble parameter in the FRW models is given by
\be
\label{Eq.hp}H\equiv \frac{\dot a}{a}=-\frac{1}{1+z}\frac{dz}{dt_{ct}},
\ee
where $a$ is the scale parameter, $dt_{ct}$ is the variation of the cosmic time due to a small change in the redshift $dz$. One can measure $H$ directly through the {\it differential age method} \citep{Jimenez2002}. We briefly review this idea below.

For any galaxy one has,
\be
\label{Eq.CA} T_{CA}(z)=T_{F}+T_{GA}(z),
\ee
which simply states that the cosmic age $T_{CA}$ at redshift $z$ equates the summation of the time for the galaxy to form, $T_{F}$, and the age of this galaxy, $T_{GA}$. The latter can be determined spectroscopically. If we could find a group of galaxies that share a uniform formation time, i.e., $T_{F}=$Const., we would then get a handle of $dt_{ct}$ by simply measuring the age difference of those galaxies:
\be
\label{Eq.dtfrw} dt_{ct}(z)=dT_{CA}=dT_{F}+dT_{GA}(z)=dT_{GA}(z).
\ee

The passively evolving galaxies can be identified by figuring out at every redshift the oldest galaxies which define the `red envelop'. In this case, one assumes the oldest galaxies formed at the same time, which is of course a natural assumption in an FRW universe (one may call it the galaxy-formation version of the cosmic Copernican principle). By observing $24$ massive galaxy clusters, determining the red envelop and differentiating the ages of the galaxies on the red envelop, \cite{Stern2010} add two new points to those of \cite{Simon2005}. The resulted dataset (hereafter the SJVKS dataset) is listed in Table \ref{tab.OHD}.

\begin{deluxetable}{cc}
\tablecolumns{2}

\tablewidth{0pt}
\tablecaption{The SJVKS dataset
\label{tab.OHD}}
\tablehead{
  \colhead{z} &
  \colhead{H(z)} \\
  \colhead{} &
  \colhead{kms$^{-1}$Mpc$^{-1}$}
}
\startdata
0.1  & $69 \pm 12$   \\
0.17 & $83 \pm 8$    \\
0.27 & $77 \pm 14$   \\
0.4  & $95 \pm 17$   \\
0.48 & $97 \pm 62$   \\
0.88 & $90 \pm 40$   \\
0.9  & $117 \pm 23 $ \\
1.3  & $168 \pm 17$  \\
1.43 & $177 \pm 18$  \\
1.53 & $140 \pm 14$  \\
1.75 & $202 \pm 40$
\enddata
\end{deluxetable}

\subsection{LTB scenario}
There are two rates of expansion in LTB models, defined respectively in Eq.\ref{Eq.Hperp} and Eq.\ref{Eq.Hparr}. It turns out that the longitudinal expansion rate $H_{\parallel}$ has the same form as Eq.\ref{Eq.hp}, hence it corresponds to the observed $H(z)$ in Table \ref{tab.OHD}. This can be seen by combining Eq.\ref{Eq.Hparr} and Eq.\ref{Eq.dzdt}, which gives:
\be
\label{Eq. HzLTB}H_{\parallel}=-\frac{1}{1+z}\frac{dz}{dt_{ct}}.
\ee

Ambiguities arise, however, from Eq.\ref{Eq.CA} when determining $dt_{ct}$. We have seen in last subsection that a same formation time of the oldest galaxies is the basic assumption in obtaining OHD. However, in LTB models where the universe has a considerable background inhomogeneity, this assumption becomes unreasonable and has to be dropped. So one question that must be answered before any data analysis is whether OHD can still be regarded as a valid dataset in the LTB scenario.

In answering this question, firstly we noticed that, albeit the overall uniform-formation-age assumption, the validity of a OHD datapoint requires a same formation time only inside the redshift bin where it is obtained (Eq.\ref{Eq.hp}-Eq.\ref{Eq.dtfrw}). This is to say, OHD data points are locally defined and obtained, and therefore can be correct even when the global density, hence the formation time of the oldest galaxies, at different redshifts varies much. Secondly, in models' side, a standard viewpoint (referred to as the Onion approximation) treats the LTB void universe as a group of thin shells structured together and that inside each of these spherical shells the matter is homogeneously distributed \citep{Biswas2007}. If the bin size for determining OHD is small enough, and the background density varies only slightly inside it, we could identify these redshift bins with the homogeneous shells above and apply the Onion approximation. Thirdly, for the SJVKS dataset, the size of each bin is between $0.1$ and $0.15$, where the first limit is so chosen that the age evolution between the two bins is larger than the error in the age determination \citep{Simon2005}. As the precision of the age determination improves, we expect an even smaller bin size.

It is by noticing the three points above that we suppose that OHD could be still used in LTB models. In order to be sure about this, however, one needs the exact knowledge about the thickness of the shell given the size of a redshift bin, as well as the steepness of the density profile at the time the oldest galaxies formed. These information will not be learned until one obtains the constraints on the model parameters. So in practice one has to first assume the Onion approximation holds, and check the validity of this assumption after the data fitting is done. The fitting result could be accepted only if the Onion approximation turns out to be valid. We will do this consistency check at the end of the next section.

\section{Constraints from the current OHD dataset}\label{sec.current}
\subsection{Likelihood}
We adopt $\chi^2$ statistics to determine the most likely values, as well as the confidence intervals for the parameter sets used in our test models. Below are the main formulas using the CGBH model as an example. For a dataset of \{$H_i^D$\} with errors \{$\sigma_{i}$\}, $\chi^2$ is defined as:
\be
\begin{split}
\label{Eq.Chi2}\chi^2(H_0, \Omega_{0}&, r_0, \Delta r)\\
=&\sum_i \frac{[H_{\parallel}^{T}(z_i; H_0, \Omega_{0}, r_0, \Delta r)-H_i^D]^2}{\sigma_{i} ^2}.
\end{split}
\ee

The likelihood function for the parameters given the dataset can be obtained via Bayes' theorem,
\be
\begin{split}
\label{Eq.pdf}\mathcal{L}(H_0, &\Omega_{0}, r_0, \Delta r|\{H_i^D\})\\
&=\frac{p(\{H_i^D\}|H_0, \Omega_{0}, r_0, \Delta r)p(H_0, \Omega_{0}, r_0, \Delta r)}{p(\{H_i^D\})}\\
&\propto p(\{H_i^D\}|H_0, \Omega_{0}, r_0, \Delta r)\\
&\propto \text{exp}\left(-\frac{\chi^2}{2}\right),
\end{split}
\ee
where the second equation comes from the fact that no priors constraints are imposed on the the data and that a uniform prior is assumed for the parameters inside their respective scanning ranges. Table \ref{tab.priors} lists the parameter range scanned in the CGBH model, for the two void models see Appendix \ref{append.svoid}.
\begin{table}
\begin{center}
\caption{Scanning priors of CGBH model parameters}
\label{tab.priors}
\vspace*{0.5cm}
\begin{tabular}{cccc}
\hline
$r_0$ & $dr$ & $H_{0}$       &              $\Omega_{0}$ \\
{\footnotesize [Gpc]} &       &{\footnotesize [km sec$^{-1}$ Mpc$^{-1}$]} &{\footnotesize  }\\
\hline
0.2-9 & 0.1-0.9      &   65-85       &              0.03-0.50 \\
\hline
\end{tabular}
\end{center}
\end{table}

\subsection{Constraints from SJVKS Data}
\begin{figure*}[!htb]
\begin{center}
\includegraphics[width=0.6\textwidth]{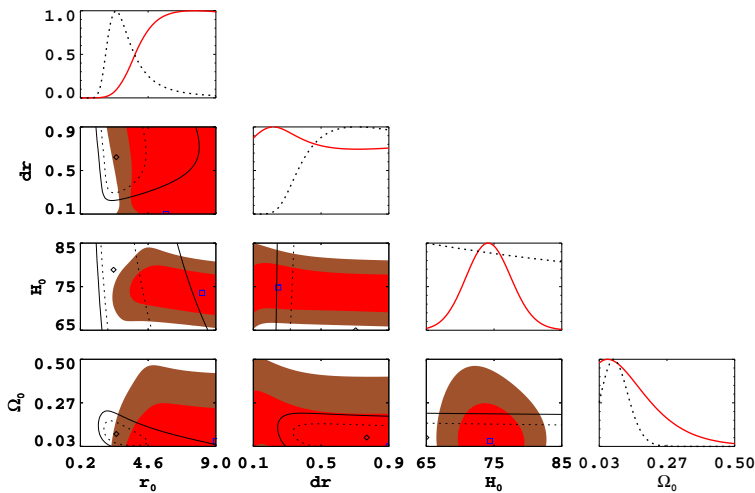}
\end{center}
\caption{Constraints on CGBH model parameters from the SJVKS dataset of OHD and the Union2 dataset of SN Ia. The red and brown filled regions correspond respectively to 1$\sigma$ and 2$\sigma$ confidence regions obtained from the SJVKS dataset of OHD. Overplotted are the 1$\sigma$ and 2$\sigma$ confidence regions obtained from the Union2 dataset of SN Ia, in solid and dotted lines, respectively.
\label{Fig.hzsn}}
\end{figure*}

We apply the $\chi^2$ analysis to the SJVKS dataset of OHD, with the underlying model being our three test models. In Figure \ref{Fig.hzsn} we show the resulted 1$\sigma$ and 2$\sigma$ marginalized likelihoods for the CGBH model. As a comparison, we also plot the likelihood contours for the latest Union2 dataset of SN Ia. Like in the FRW case, $H_0$ in LTB models is treated as a nuisance parameter for SN Ia data (see Appendix \ref{append.nuisance} for a discussion on this point). Table \ref{tab.bestfit} lists the $1-$D best-fit parameters of the CGBH model. Since the exact fitting result of a particular model is of only secondary importance, we put the other two models' results in Appendix \ref{append.svoid} just for reference.
\begin{table}
\begin{center}
\caption{Best-fit CGBH model parameters}
\label{tab.bestfit}
\vspace*{0.5cm}
\begin{tabular}{ccccc}
\hline
 \textbf{CGBH} & $r_0$ & $dr$ & $H_{0}$       &              $\Omega_{0}$ \\
 & {\footnotesize [Gpc]} &       &{\footnotesize [km sec$^{-1}$ Mpc$^{-1}$]} & \\
\hline
OHD & 7.56 & 0.21     &   74       &              0.058 \\ 
\hline
SN & 2.53 & 0.70      &   /       &              0.078 \\ 
\hline
OHD+SN & 3.60 & 0.85      &   74       &              0.049 \\ 
\hline
\end{tabular}
\end{center}
\end{table}

\begin{figure}[!htb]
\begin{center}
\includegraphics[width=\columnwidth]{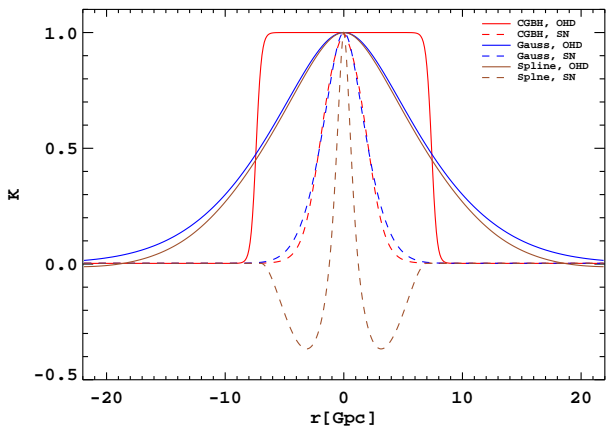}
\end{center}
\caption{Curvature profiles of the best-fit models from the SJVKS dataset of OHD and the Union2 dataset of SN Ia. OHD prefers voids with curvature changing slow near the origin; while the best-fit voids obtained from SN Ia are more cuspy.  \label{Fig.voidk}}
\end{figure}
Neither dataset gives satisfactory constraints. Nevertheless, several instructive clues can be observed from Figure \ref{Fig.hzsn}: 1)OHD prefers a void profile that is considerably broader than that from SN Ia; 2)Both SN Ia and OHD prefer a void that is almost depleted at the origin; 3)Aside from the nuisance parameter $H_0$, the SN Ia Union2 dataset is generally more constraining than the SJVKS dataset of OHD.

Note that the 1-D likelihood of $dr$ from OHD is far from being Gaussian and the best-fit value is seen not so well-defined. This in turn weakens the first point of the conclusions above. However, we find that the same three conclusions hold for the two smooth void models too, despite their totally different parameterisation. This might be a sign that the differences are real and are model-insensitive. We are particularly interested in the first point: if the future data of OHD and SN Ia both hold their different preferences of the void profiles, the smaller confidence regions they give would finally detach with each other. This would then be a sign that the void model is incorrect and should be excluded. In Section \ref{sec.forecast} we will give a rough prediction on this based on simulated OHD data.

\subsection{Validity check of the Onion approximation}
\begin{figure}[!htb]
\includegraphics[width=\columnwidth]{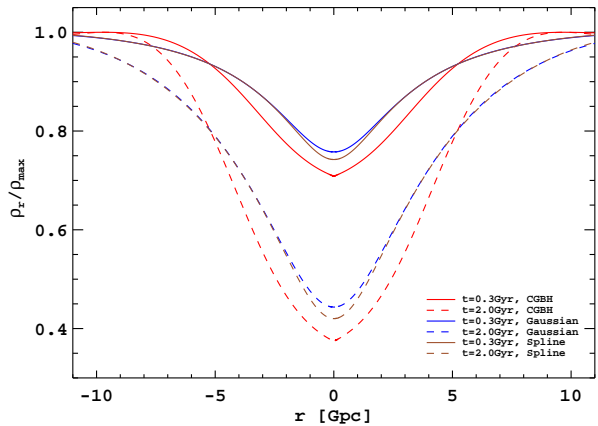}
\caption{Density profiles at different cosmic time. As the voids get deeper, the background density become more inhomogeneous. \label{Fig.rho_tslice}}
\end{figure}

\begin{figure}[!htb]
\includegraphics[width=\columnwidth]{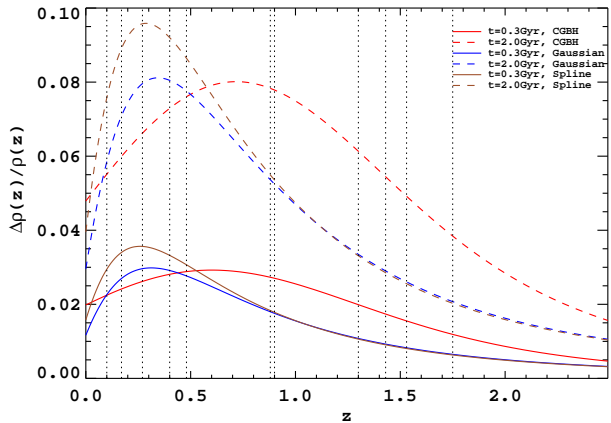}
\caption{Relative density variation inside a redshift bin of $0.15$ as a function of the redshift. $\Delta\rho(z)\equiv\rho(z+0.15)-\rho(z)$, where $\rho(z)\equiv\rho(r(z))$ is the density value at $r$ on a given time slice, and $r$ is converted from $z$ using the best-fit models. The dotted vertical lines indicate the data points of the SJVKS dataset, so one can tell approximately the relative background density variation inside the bins where each data point is obtained. \label{Fig.drhodz}}
\end{figure}

Now that we have obtained the best-fit void models, we can use them to check if the Onion approximation we have made indeed holds, i.e., if the universe can be treated as homogeneous inside the redshift bins where the OHD data points are obtained. We take the best-fit parameters given by the SJVKS dataset and show in Figure \ref{Fig.rho_tslice} the density profile at different cosmic times. The inhomogeneity is seen to become more significant and the void gets deeper as the universe evolves, because of a higher expansion rate inside the void. In order to see more clearly how the density varies inside a small redshift bin of which we are more concerned, we plot in Figure \ref{Fig.drhodz} the relative density variation inside a $\Delta z=0.15$ bin. More specifically, we calculate on a given time slice $\Delta\rho(r(z))/\rho(r(z))$, where $\rho(r(z))$ is the density value at $r$, with $r$ converted from $z$ using the best-fit parameters. As stated in the last section, each OHD data point is obtained using galaxies that are in a redshift bin no larger than $0.15$, so $\Delta z=0.15$ is a conservative choice for our purpose. At $T_{CA}=0.3$Gyr this relative change of the background density are globally less than $4\%$ (see Figure \ref{Fig.drhodz}). If the oldest galaxies were formed around this time, which is not unreasonable for structures to form, the Onion approximation is a reasonable one. On the other hand, if these galaxies formed too late, the influence of the background density variation on the structure formation may become too great to ignore.

To determine this formation time, Eq.\ref{Eq.CA} is used and from which we have $T_F=T_{CA}(0)-T_{GA}(0)$. The age of the oldest local galaxies, $T_{GA}(0)$, can be approximated by taking the Y-intercept of the red envelop\footnote{As we don't expect a homogeneous age of all the first galaxies in the LTB models, the red envelop does not have the simple relation with the cosmic time as it does in the FRW models. But the ages of the galaxies on the red envelop can still serve as a lower bound of the ages of the universe at the corresponding redshifts.} on Figure 11 of \cite{Stern2010}. We find however, the outcome depends strongly on the models used to fit the galaxy spectra, which could sometimes yield a $T_{GA}(0)$ greater than 14Gyr. This number would defy all our best-fit void models, as well as the latest $\Lambda$CDM model result, $13.79$Gyr \citep{Komatsu2011}, because $T_F$ would then be negative. So finally, we could {\it not} know for sure whether the Onion approximation is valid or not. Only qualitatively could we say that the oldest galaxies observed seem to be formed very early, so very likely the SJVKS dataset is valid. More precise observations are needed to determine $T_{GA}(0)$, hence $T_F$. As we need to use the best-fit models to generate simulated data, we leave this as an open problem and {\it assume} from now on that the Onion approximation is justified.

\section{Comparison between Best-fit Voids and flat $\Lambda$CDM}\label{sec.vslcdm}
\subsection{AIC$_c$ Analysis}
\begin{figure}[!htb]
\begin{center}
\includegraphics[width=\columnwidth]{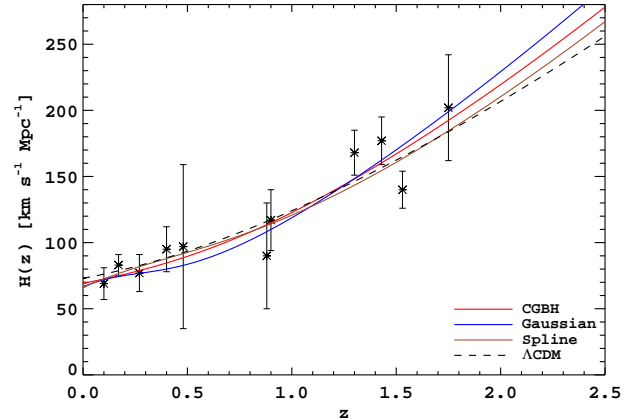}
\end{center}
\caption{Hubble parameter as a function of the redshift. The prediction from the best-fit CGBH model based on the SJVKS dataset is plotted in red solid line and that from the {\flcdm} model is plotted in blue dashed line. Also plotted are the SJVKS dataset.
\label{Fig.hzcomp}}
\end{figure}
The theoretical prediction from different models at their best-fit parameters, as well as the observational data of OHD and SN Ia, are plotted in Fig.\ref{Fig.hzcomp}, and Fig.\ref{Fig.hnorm}, respectively. The current observation of these two probes are not able to exclude either. To determine which model is preferred by the data, a useful tool is the Akaike Information Criterion (AIC, \cite{Akaike1974}). The AIC for a model is
\be
\text{AIC}=2k-2\text{ln}\mathcal{L}_M,
\ee
where $k$ is the number of the parameters, and $\mathcal{L}_M$ is the maximized likelihood function. The preferred model is the one with the minimum AIC value, so not only does AIC reward a better goodness of fit, it also punishes a larger number of parameters.

For OHD however, the number of the data points, $n$, is comparable to $k$, and the probability of overfitting becomes considerable. A second-order correction of AIC, AIC$_c$, should be used in such cases \citep{Sugiura1978, Hurvich1989, Burnham2002}:
\be
\text{AIC}_c=AIC+\frac{2k(2k+1)}{n-k-1}.
\ee

AIC$_c$ punishes the extra parameters more harshly, and it converges to AIC when $n$ becomes large. Note that only the difference between the AIC$_c$ is meaningful, so it is convenient to omit the constant in ln$\mathcal{L}_M$, which is the same for the same dataset, and express AIC$_c$ in $\chi^2$:
\be
\label{Eq.aicc}\text{AIC}_c=2k+\chi^2_\text{min}+\frac{2k(k+1)}{n-k-1}.
\ee

We list the AIC$_c$ values for our LTB test models and {\flcdm} in Table \ref{tab.aic}. We discuss this result again using the CGBH model as an example. Although having an AIC$_c$ calculated from SN data that is fairly close to that of the {\flcdm}, the CGBH model is considerably less supported by OHD. This is not to say that the CGBH model fits OHD worse than {\flcdm}. On the contrary, the minimum $\chi^2$ from OHD dataset are $\chi^2_\text{min}($CGBH$)=6.97$ and $\chi^2_\text{min}(\Lambda\text{CDM})=7.66$, respectively, i.e., the former fits the data better. Therefore, of the three terms in Eq.\ref{Eq.aicc}, the third contributes the most to the difference in AIC$_c$. As this term will get smaller as the dataset enlarges, the performance of CGBH in the AIC$_c$ test against OHD is expected to get better in the future.
\begin{table}
\begin{center}
\caption{AIC$_c$ value for different models}
\label{tab.aic}
\vspace*{0.5cm}
\begin{tabular}{ccccc}
\hline
& CGBH & Gaussian & Spline & {\flcdm} \\
\hline
OHD & 21.7 & 16.4 & 29.1 &13.2\\
\hline
SN & 537.9 & 536.9 & 539.5 &535.1\\
\hline
\end{tabular}
\end{center}
\end{table}

\subsection{Effective Quantities}
To better visualize the evolution of the observable LTB universe, it is usually convenient to construct some `effective' parameters in a way originally used in FRW cosmology. The evolution of these parameters basically tells us what we would observe if we assume an FRW universe, while the real universe is, say, an LTB void. The best-fit model mentioned below will be the one determined from OHD and SN Ia combined.

\begin{figure}[!htb]
\begin{center}
\includegraphics[width=\columnwidth]{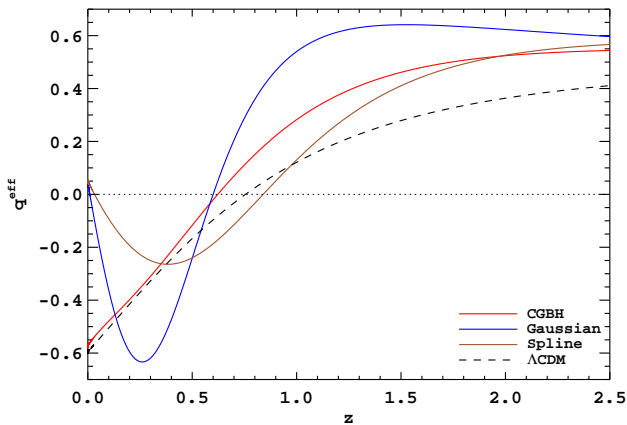}
\end{center}
\caption{The effective deceleration parameter as a function of the redshift. The red solid line corresponds to the best-fit CGBH model, the blue solid line the Gaussian void, the brown solid line the Spline void, and the black dashed line the best-fit \flcdm.
 \label{Fig.effacc}}
\end{figure}

The first such parameter we discuss in this subsection is the effective deceleration parameter, defined to be
\be
q^\text{eff}=-1+\frac{d\text{ln}H_{\parallel}}{d\text{ln}(1+z)},
\ee
which in FRW cosmology would be just the familiar deceleration parameter $q$. In Figure \ref{Fig.effacc} we show $q^\text{eff}$ derived from the best-fit void models, compared against the deceleration parameter calculated from the best-fit {\flcdm} model. The void models are seen to mimic observationally an universe that experiences an acceleration period between $z=[0, 1]$.
\begin{figure}[!htb]
\begin{center}
\includegraphics[width=\columnwidth]{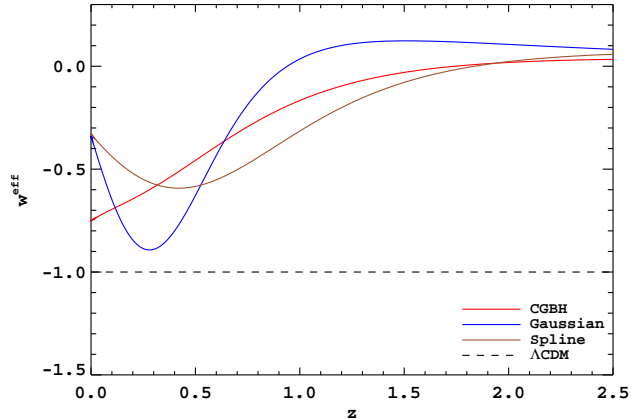}
\end{center}
\caption{The effective Dark Energy equation of state as a function of the redshift. The red solid line corresponds to the best-fit CGBH model, the blue solid line the Gaussian void, the brown solid line the Spline void,and the blue dashed line the best-fit \flcdm.
 \label{Fig.effw}}
\end{figure}

The second effective quantity is the effective equation of state (EoS) of the dark energy. Before the existence of the dark energy is unambiguously verified, one should always bear in mind that the dynamical properties of the dark energy people try to figure out today may be really an effective effect of, say, a background inhomogeneity. This effective parameter can be constructed again from $H_\parallel$:
\be
w^\text{eff}=-1+\frac{1}{3}\frac{d\text{ln}\mathcal{Q}}{d\text{ln}(1+z)},
\ee
where
\be
\mathcal{Q}=\frac{H_\parallel(z)^2}{H_\parallel(0)^2}-\Omega_0(1+z)^3.
\ee

The evolution of this effective EoS is plotted in Figure \ref{Fig.effw}. Contrary to FLSC, we did not find a $-1$ crossing of the effective EoS in any of our test voids, which might be partly due to the more recent datasets of both SN Ia and OHD we have used.

\section{Forecast of constraints from the future OHD dataset}\label{sec.forecast}
We have seen in the last section that 1)the two-dimensional confidence regions of the parameters from OHD and SN Ia are overlapped (consistent), 2)the two observational probes prefer characteristically different void profiles. In this section, we explore the situation in the future with Monte Carlo simulated OHD data. The first subsection is devoted to explaining the essential ingredients of our simulation.

To facilitate the comparison between SN Ia and OHD, we introduce for each void model a characteristic parameter plane, which we refer to as CPP for brevity. The constraining power of one observational probe is quantified via the area of the confidence region on this specific parameter plane. As we will find later, this is directly related to the definition of the figure of merit. Take CGBH model for example, the two parameters $r_0$ and $dr$ contains most of the information one would like to know about the void profile, so $r_0-dr$ plane is chosen as its CPP. The CPPs of the Gaussian void and Spline void are not so obvious. In fact for the Gaussian void model, a better way might be to compare the length of the $1$-Dim parameter confidence interval, and the Spline model the volume of the $3$-Dim ($\Omega_{r2}, r2, r3$) parameter confidence region. Here for the sake of consistency, we look for the best parameter pair and choose the CPP for the Gaussian and the Spline void models to be $r_0-\Omega_0$ and $r_2-\Omega_{r2}$, respectively.

Based on the simulated data two questions are investigated, both of which are related to the CPP chosen above. In the following we describe them in turn. Firstly, it is possible that when more OHD data points are obtained, they become more consistent with the SN Ia data. On the CPP, this would mean that the $1\sigma$ and $2\sigma$ confidence regions obtained from OHD become more overlapped with those from SN Ia. One would like to compare in this case the constraining power of the two probes. The first question (Question-I) is therefore raised, that when the area of the OHD's confidence regions on the CPP become smaller than that of SN Ia. The standard way to do this type of comparison is via the Figure of Merit (FoM). We calculate FoMs for our simulated datasets in the second subsection.

Considering the different preference in the void profiles of OHD and SN Ia, the second question (Question-II) one would like to know through OHD data simulation is how many OHD data points are needed for the $1\sigma$ confidence region on the CPP totally detached from that of the SN Ia, if the future OHD data points hold the same void-preference as the current ones. This will be the content of the third subsection.

\subsection{Monte Carlo Simulation}
\textbf{The fiducial function of $H(z)$.}
This function sets the expectation value of the simulated OHD data. For the purpose of Question-II, where the future data points are assumed to have the same void preference as the current ones, a straightforward fiducial function of $H(z)$ comes from the best-fit void models obtained from the SJVKS dataset of OHD.

In dealing with Question-I, on the other hand, where we assume the future OHD and SN Ia `converge' in their best-fit void profiles, hence the model parameters, one natural choice is again to use the best-fit void models; only this time the model parameters are given by OHD and SN Ia {\it combined}. The problem is that the constraints from the current datasets are far from tight, as we have seen in Figure \ref{Fig.hzsn}; the best-fit parameters are not Gaussian in many cases, hence not quite well-defined. The ({\flcdm}) model, on the other hand, is consistent with and well-constrained by all current observations, and therefore generalize the best knowledge we have about the observational universe; a fiducial model based on this model would seem to be a proper choice. However, {\flcdm}, as do any other FRW models, differs characteristically from the void models. Using {\flcdm} as the fiducial model, one risks introducing a systematic error, which would in turn cause a shift and/or a shape change in the parameter confidence region \citep{Kim2004}, hence a bias in the figure of merit (see the next subsection). Noticing the merits and demerits of each of the two choices, we obtain the center values of the simulated dataset from a mixture of them \citep{Robert2004}. More specifically, statistically half of the datapoints are drawn from the {\flcdm}, with $\Omega_m=0.27$, $H_0=73$, which is consistent with the 7-year Wilkinson Microwave Anisotropy Probe (WMAP) and $H_0$ observation \citep{Komatsu2011, Riess2011}; and in generating the other half of the data, the void model with the combined best-fit parameters is used as the fiducial function.

The discussion of the following two ingredients of Monte Carlo simulation should be the same when applied to Question-I and Question-II. Where the fiducial function is used for illustration, we use the best-fit void model (combined).

\begin{figure}[!htb]
\begin{center}
\includegraphics[width=\columnwidth]{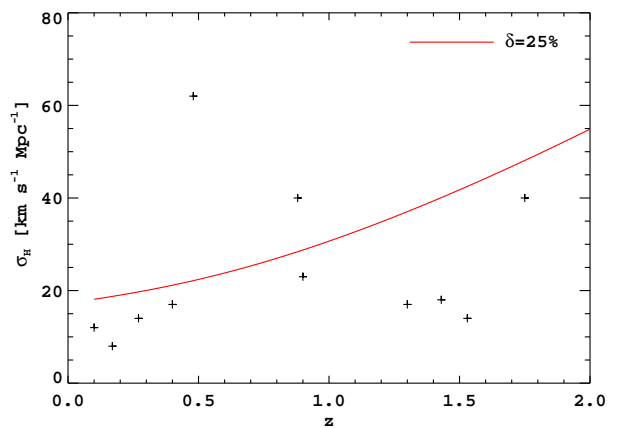}
\end{center}
\caption{The $H_i\delta$ line with $\delta=25\%$ forms an envelop of the errors of SJVKS dataset. Only two data points at $z=0.48, 0.88$ are seen to be beyond the envelop.
 \label{Fig.envelop}}
\end{figure}

\textbf{The uncertainty model.} This model generalizes the statistical and systematic uncertainties about a future observation. There is, however, no such specifications for future OHD survey, to the authors' knowledge. To work around this difficulty, \cite{Ma2011} examined the SJVKS dataset and proposed a simple phenomenological uncertainty model. Here, in order to gain more control on the simulations, we adopt an even simpler strategy. We introduce a percentage, $\delta$, to describe the maximum uncertainty for each data point, namely, for a given fiducial value $H_i$, the expectation of the uncertainty $\sigma_i$ is taken randomly from the interval [-$H_i\delta$, $H_i\delta$]. This uncertainty model is surely over-simplified: not all errors in the observations are redshift-dependent. Nonetheless, as shown in Figure \ref{Fig.envelop}, a $\delta=25\%$ error line does act as an envelop of the current observational uncertainties, except at $z=0.48, 0.88$, where the data points have extraordinary errors \citep{Stern2010a}.

\begin{figure}[!htb]
\begin{center}
\includegraphics[width=\columnwidth]{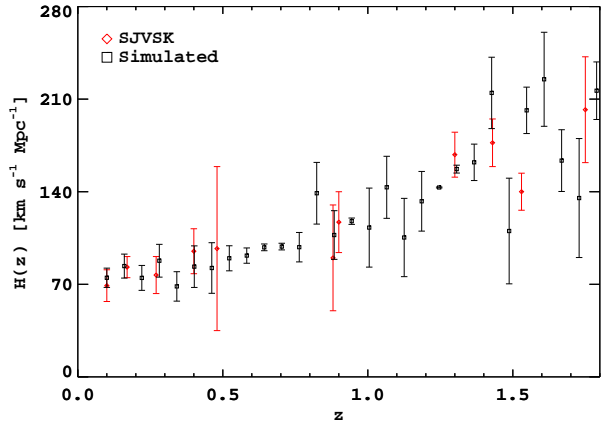}
\end{center}
\caption{One snapshot realization of the OHD dataset, with the uncertainty level $\delta=25\%$. The SJVKS dataset (diamond with red errorbar) are plotted for comparison.
 \label{Fig.hzsimu}}
\end{figure}

Once $\sigma_i$ is obtained, the mid-point of the observational data will then be generated from a Gaussian distribution $\mathcal{N}(H_i, \sigma_i^2)$. One such sample is shown, against the SJVKS dataset, in Figure \ref{Fig.hzsimu}.

\textbf{Sample size and survey depth.} The survey depth and the sample size are another two important aspects of an observation that influence much the parameter determination. We denote them as $z_\text{max}$ and $N_s$, respectively. Therefore we have in total three control parameters to describe a possible future observation (or to generate one sample of data points): \{$\delta, z_\text{max}, N_s$\}. In practice, $\delta$ takes three values: $25\%$ (current precision), $15\%$ (according to a prediction of OHD from Atacama Cosmology Telescope, see \cite{Simon2005}), $3\%$ (now can be achieved at $z\sim0.42$, see \cite{Crawford2010}); $z_\text{max}$ takes also three values: $1.5$, $2$, $2.5$, and $N_s$ takes four: $64, 32, 16, 8$.

\subsection{Figure of Merit}
\begin{table}
\begin{center}
\caption{FoM of different void models}
\label{tab.fom}
\vspace*{0.5cm}
\begin{tabular}{ccccc}
\hline
& CGBH & Gaussian & Spline\\
\hline
OHD & 0.183 & 0.107 & 2.508\\
\hline
SN & 0.272 & 0.712 & 3.707\\
\hline
\end{tabular}
\end{center}
\end{table}
We define the Figure of Merit (FoM) to be the reciprocal area enclosed by the $2\sigma$ marginalized likelihood contour on the CPP of each of our void models. Higher FoM would then mean tighter, hence better constraint. The current values of FoM from OHD and SN Ia for different void models are shown in Table \ref{tab.fom}, the values in the OHD row are seen to be less than their counterparts in the SN row. Question-I could be translated then to the `FoM language' as: when will the future OHD dataset give out a FoM larger than that of the SN Ia Union2 dataset?

To calculate the FoMs of the simulated datasets with fixed $z_\text{max}$ and $\delta$, three steps are taken in turn: 1) generating a sample of $64$ OHD data points that are evenly spaced in the redshift range $[0.1, z_\text{max}]$; 2) calculating the FoM based on this sample; 3) Extracting subsamples with $N_s=32$, $N_s=16$, $N_s=8$ and calculating their FoMs in turn. Repeat these three steps for, say, $100$ times, and we get for each $N_s$ an array of $100$ FoMs, the median of which is taken as the typical FoM. Then we iterate the whole process for different \{$z_\text{max}, \delta$\} and finally get a 3-D matrix, with the element being the typical FoM at each \{$z_\text{max}, \delta, N_s$\}.

\begin{figure*}[!htb]
\begin{center}
\includegraphics[width=0.48\textwidth]{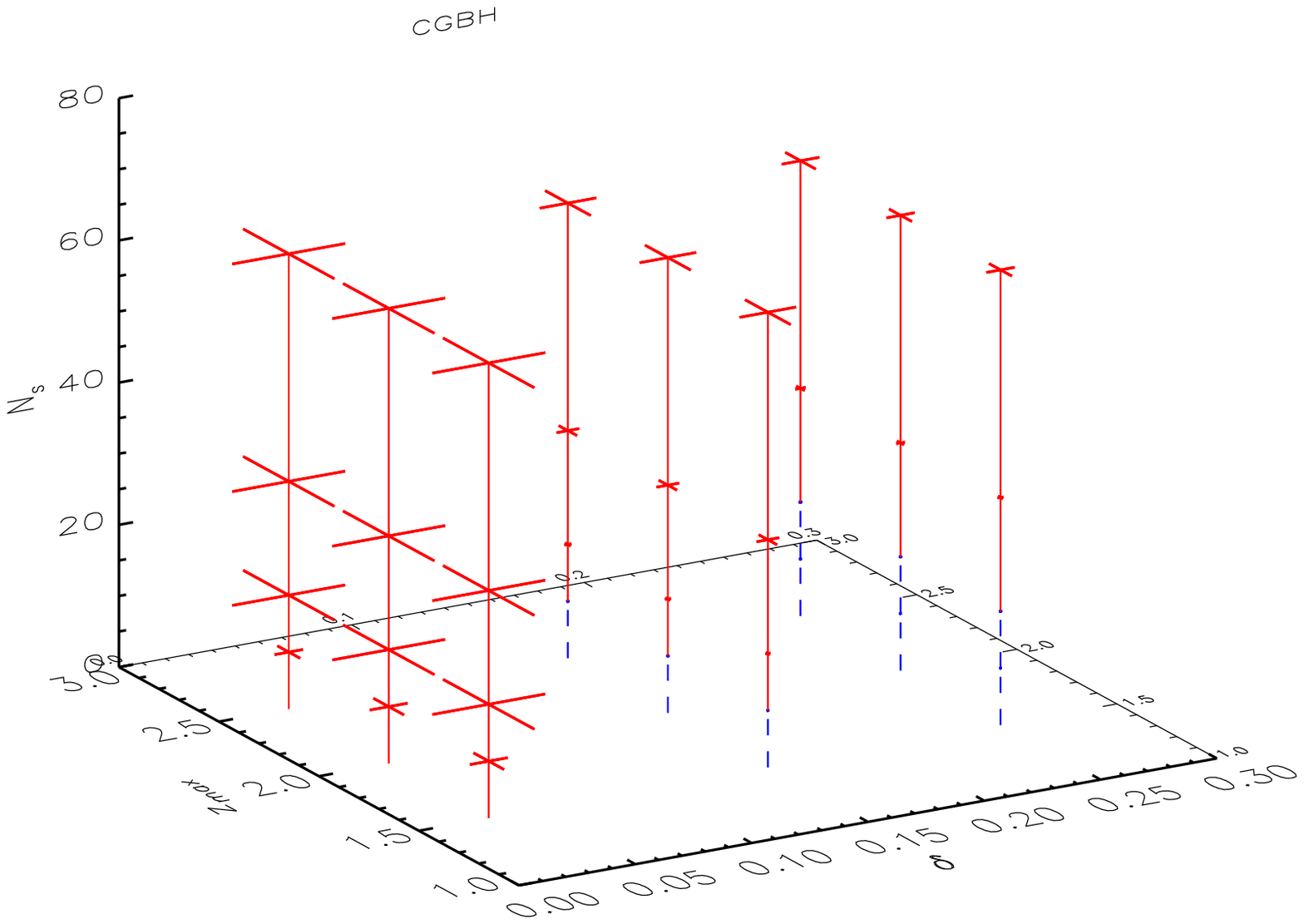}
\includegraphics[width=0.48\textwidth]{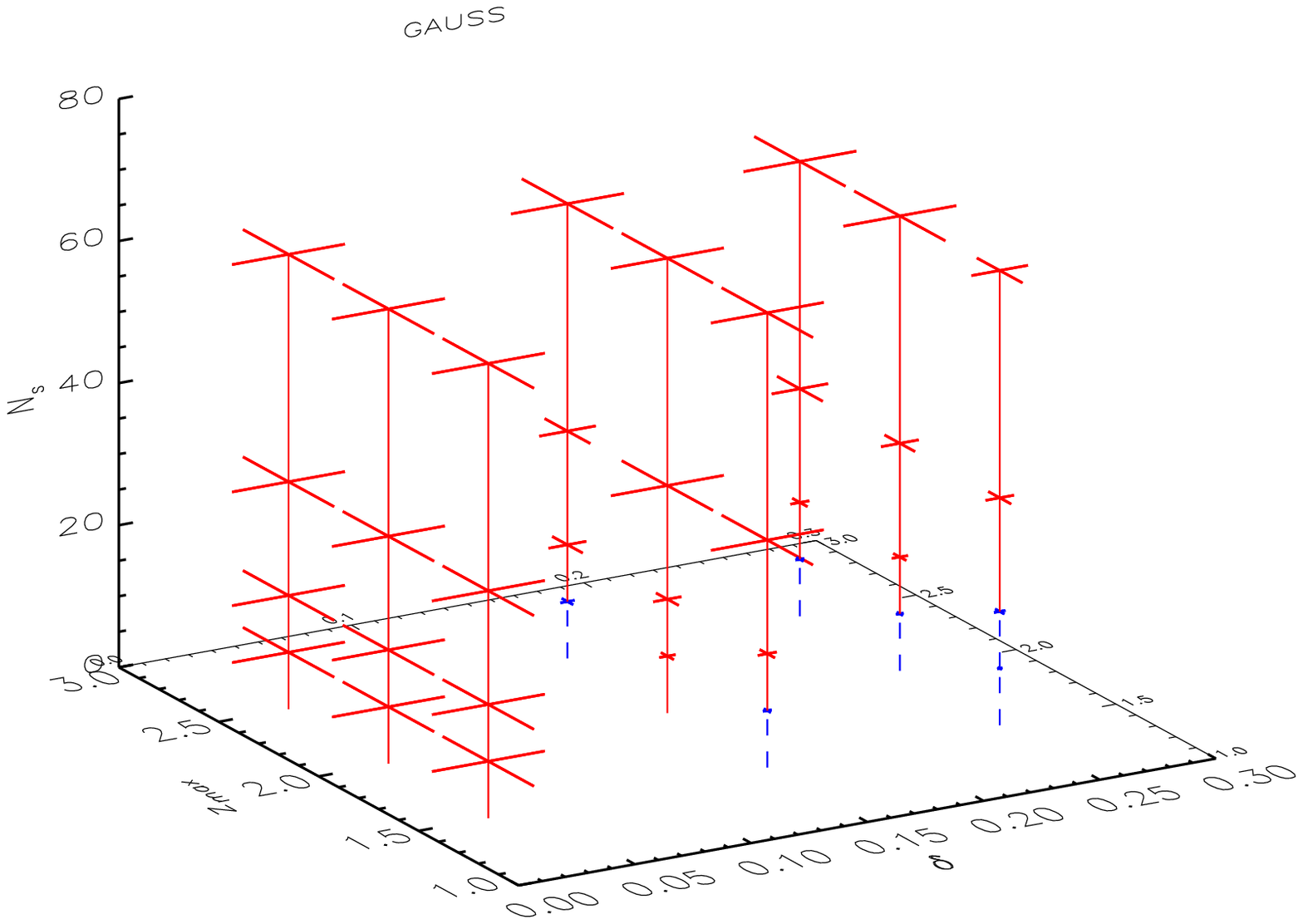}
\includegraphics[width=0.48\textwidth]{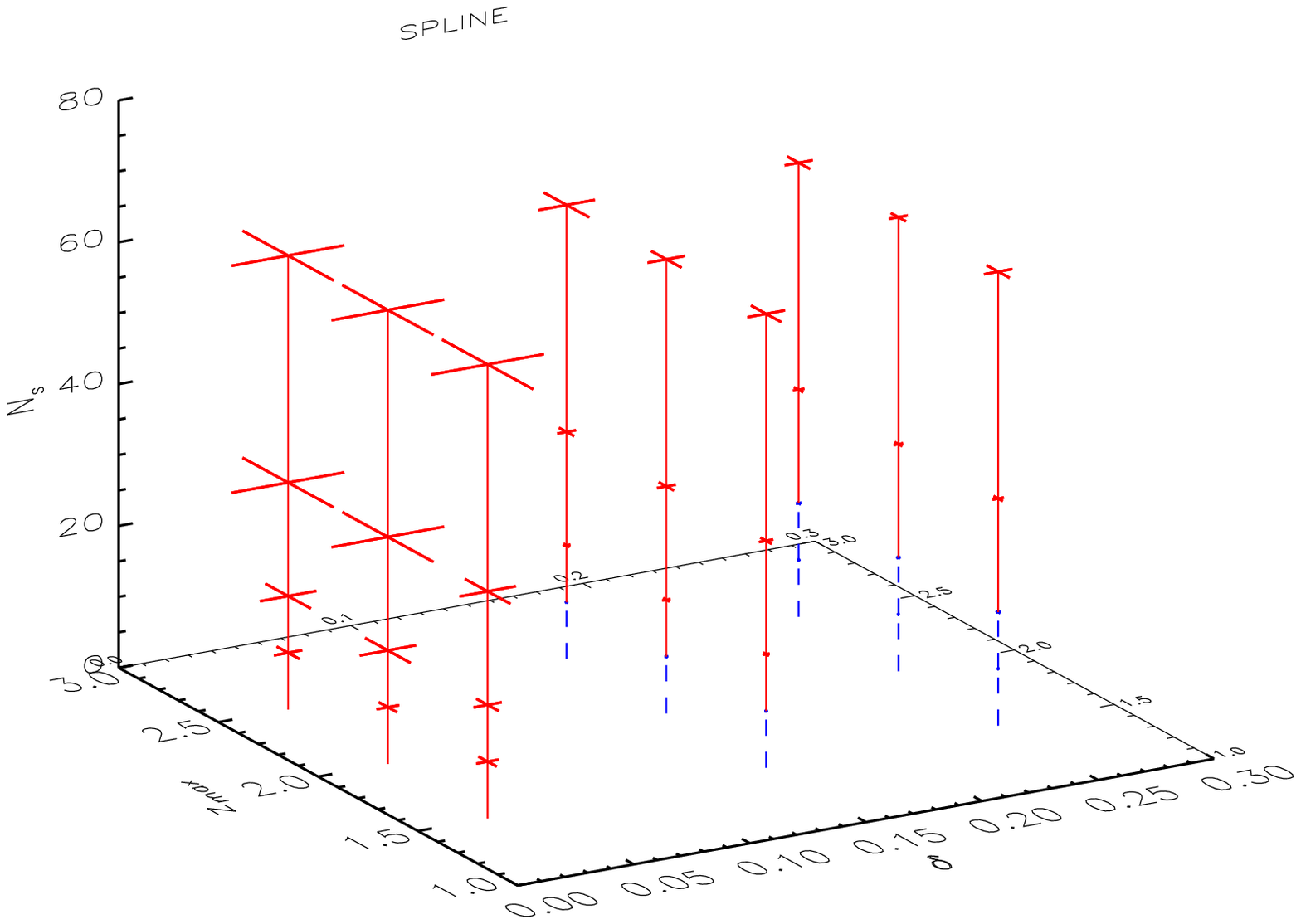}
\end{center}
\caption{The Figure of Merit (FoM) in the control parameter space. In these right-handed coordinate systems, the x-axes are the error level $\delta$, the y-axes $z_\text{max}$ and the z-axes the size of the datasets $N_s$. Length of the plus symbol ($+$) are proportional to the values of the FoM. The symbol is plotted in red, and the line connected to it from below is solid, if the FoM value is greater than that of the Union2 dataset of SN Ia; blue and dashed otherwise.
\label{Fig.FoM}}
\end{figure*}

The result is shown in Figure \ref{Fig.FoM}. The size of the `$+$' symbol is proportional to the value of the FoM, which, if greater than that of the Union2 dataset of SN Ia, is plotted in red, otherwise blue.

Three void models are seen to share several very similar features, both qualitatively and quantitatively. Firstly, as one may intuitively assume, the FoM increases with the sample size $N_s$. At $\delta=25\%$, an uncertainty level similar to SJVKS, a sample of $32$ data points will surpass SN Union2 for all choices of $z_\text{max}$.

The second point is not surprising either, that lowering the uncertainty level $\delta$ will increase the FoM. At $\delta=15\%$, a sample of $16$ OHD data points are sufficient to surpass SN Union2 for all choices of $z_\text{max}$. If we could achieve an overall $\delta=3\%$ precision, $N_s=8$ is already good enough to give a FoM larger than that of SN Union2.

The third point deserves a bit more discussion. One might expect that a deeper survey will tighten the constraints. This is not always true, however, judging from the value of the FoMs. For example, when $N_s=16$ and $\delta=25\%$, the FoM of OHD with $z_\text{max}=1.5, 2.0, 2.5$ are $0.209, 0.203, 0.234$, respectively. That is, higher $z_\text{max}$ might lead to a lower FoM. This phenomena exist in all three void models. One possible reason for this to happen is that, for a given $N_s$, especially when $N_s$ is small, a higher $z_\text{max}$ leads to a looser redshift coverage, which in turn leads to more freedom for the theoretical curve to vary between the adjacent data points. The extra freedom in the theoretical curve is realized of course through the extra freedom of the model parameters, hence a lower FoM.

\subsection{$1\sigma$ confidence region overlap}
As discussed above, Question-II is motivated by the difference in the best-fit models obtained from the SJVKS dataset of OHD, and that from Union2 dataset of SN Ia. The logic is that if this difference holds for the future data, then one could finally tell the inconsistency from the model parameter confidence regions obtained from OHD and SN Ia, for instance, their $1\sigma$ confidence regions on the CPP will separate.

To simulate this, we generate new OHD data using the SJVKS-induced best-fit void models as the fiducial function. We `mark' at first those parameters encompassed by the SN Union2-induced $1\sigma$ contour on the CPP. Then for each simulated OHD dataset, we determine its likelihood and count the number of the marked parameters lying inside its $1\sigma$ confidence region, $\bm{N}_o$. For each simulation control parameter combination, we generate 100 samples, obtain 100 $\bm{N}_o$s and log the median value. This median number then reflects the typical overlapping condition for a given \{$N_s, \delta, z_\text{max}$\} combination, i.e., if it equals zero then very likely one could see the separation of the OHD- and SN Ia-induced $1\sigma$ confidence regions on the CPP.

The results from different void models are not so homogeneous as in the last subsection. The Gaussian void model is the easiest for the inconsistency to emerge: most of the control parameter combinations except those with $N_s=8, \delta=25\%$ have a median $\bm{N}_o$ equal to 0. For CGBH model, only at the largest dataset size ($N_s=64$), the highest redshift reach ($z_\text{max}=2.5$ and the highest precision ($\delta=3\%$) is the median value of $\bm{N}_o$ zero. In the case of the Spline void, $\bm{N}_o$ is nonzero for all the control parameter combinations, that is, from $r_2-\Omega_{r2}$ plane at least SN Ia and the OHD are always consistent. Therefore, the prediction about when the OHD observation is powerful enough that, when combined with SN Union2 dataset, can help rule out the void models, depends on the specific void model. This is our answer to Question-II.

\section{Conclusions}\label{sec.conclusion}
We have discussed in this work the use of OHD, both the present and in the future, as a probe to constrain the LTB models.

We have firstly showed, that of the two expansion rates involved in LTB models, the one in the radial direction corresponds to the observable in OHD, which are obtained from the passively evolving galaxies. The latest OHD compilation, also the one discussed in this work, is the SJVKS dataset updated and compiled in \cite{Stern2010}. A subtlety is then discussed on using OHD in the LTB scenario: on one hand, a synchronized oldest-galaxies forming time through out the universe is the basic assumption behind the OHD observation; on the other hand, LTB models have large scale background inhomogeneities, and there is no good reason to believe the first galaxies formed simultaneously all over the universe.

The problem is severe because if the dataset itself is problematic, surely it can not be used to constrain any model. We have examined the method in obtaining OHD and found that only the homogeneity inside the redshift bin ($\leq0.15$), where each data point is obtained, is demanded for the data to be valid. This is not a unreasonable assumption for the LTB models. In the standard Onion approximation, the LTB models are approximated as being composed by different spherical layers, inside which the matter is treated as homogeneously distributed and the dynamics are FRW-like. However, whether a redshift bin can be viewed as one such layer should be checked given the model parameters. We argue therefore a complete parameter constraining analysis should have a consistency check procedure after the standard $\chi^2$ analysis, and we illustrated this process on our test models with the SJVKS data at the end of Section \ref{sec.current}. The result is inconclusive due to the large uncertainties in the absolute ages of the oldest galaxies.

The constraints obtained from the likelihood analysis is indicating. SN Ia performs generically better than OHD judging from the likelihood confidence regions, but the two observational probes both favor the voids with the center largely depleted of matter. Besides, and more remarkable still, although the $1\sigma$ confidence regions from OHD overlap with that from the SN Ia data in all of the test models, the former favors the broad and flat voids, while the latter prefers voids that are smaller and more steep around the center.

In Section \ref{sec.vslcdm} we compared the best-fit void models with the best-fit \flcdm, firstly through the corrected Akaike Information Criterion (AIC$_c$) analysis. The latter, according to the AIC$_c$ value, is better than any of our void models. Indeed, especially for the complicated models, the small size of the SJVKS dataset of OHD greatly degrade their performance in the AIC$_c$ analysis.

The dynamical properties are then compared, in terms of the effective deceleration parameter ($q^\text{eff}$) and the effective equation of state (EoS) of dark energy ($w^\text{eff}$). $q^\text{eff}$ is seen to enter the negative region recently ($z<1$) for all the void models, mimicking the recent acceleration of the universe. $w^\text{eff}$ of the void models are negative today, mimicking the dark energy's repulsive power.

This general consistence and the characteristic difference in the best-fit voids of SN Ia and OHD, obtained in Section \ref{sec.current}, are taken further by using the Monte Carlo simulated data in Section \ref{sec.forecast}. More specifically, we use the simulated data to predict when OHD become more constraining than the Union2 dataset of the SN Ia (Question-I), and when the two probes become inconsistent with each other (Question-II), respectively. We make use of the Figure of Merit (FoM) analysis for Question-I, and the answer turned out to be surprisingly homogeneous for the three models: $N_s=32$ is needed for the FoM of OHD to be greater than that of Union2 dataset, when $\delta=25\%$; $N_s=16$ when $\delta=15\%$; and $N_s=8$ when $\delta=3\%$; increasing $z_\text{max}$ does not necessarily increase the constraining power.

The answer to Question-II is not so simple and is only semi-quantitative: to observe an inconsistency for the Gaussian void model, a dataset with slightly more data points than SJVKS ($N_s=16$), and at the current error level ($\delta=25\%$) will do the job; but only when $N_s$, $\delta$, $z_\text{max}$ are all at their most demanding level ($64, 3\%, 2.5$, respectively) does the inconsistency show up; in the model with the most parameters, the Spline void model, the simulated OHD and SN Ia are consistent for all the possible \{$N_s, \delta, z_\text{max}$\} combinations.

\acknowledgements
We are grateful to the anonymous referee for valuable comments that help greatly improved the paper. Hao Wang thanks Nabila Aghanim, Marian Douspis and Mathieu Langer for discussions on the Non-Copernican cosmology, which initiated this work. We thank Ma Cong for useful discussions on the subject of sample mixing. This work was supported by the National Science Foundation of China (Grants No. 11173006), the Ministry of Science and Technology National Basic Science program (project 973) under grant No. 2012CB821804, and the Fundamental Research Funds for the Central Universities.

\appendix
\section{$H_0$ as a normalization factor}\label{append.nuisance}
\begin{figure}[!htb]
\begin{center}
\includegraphics[width=0.32\textwidth]{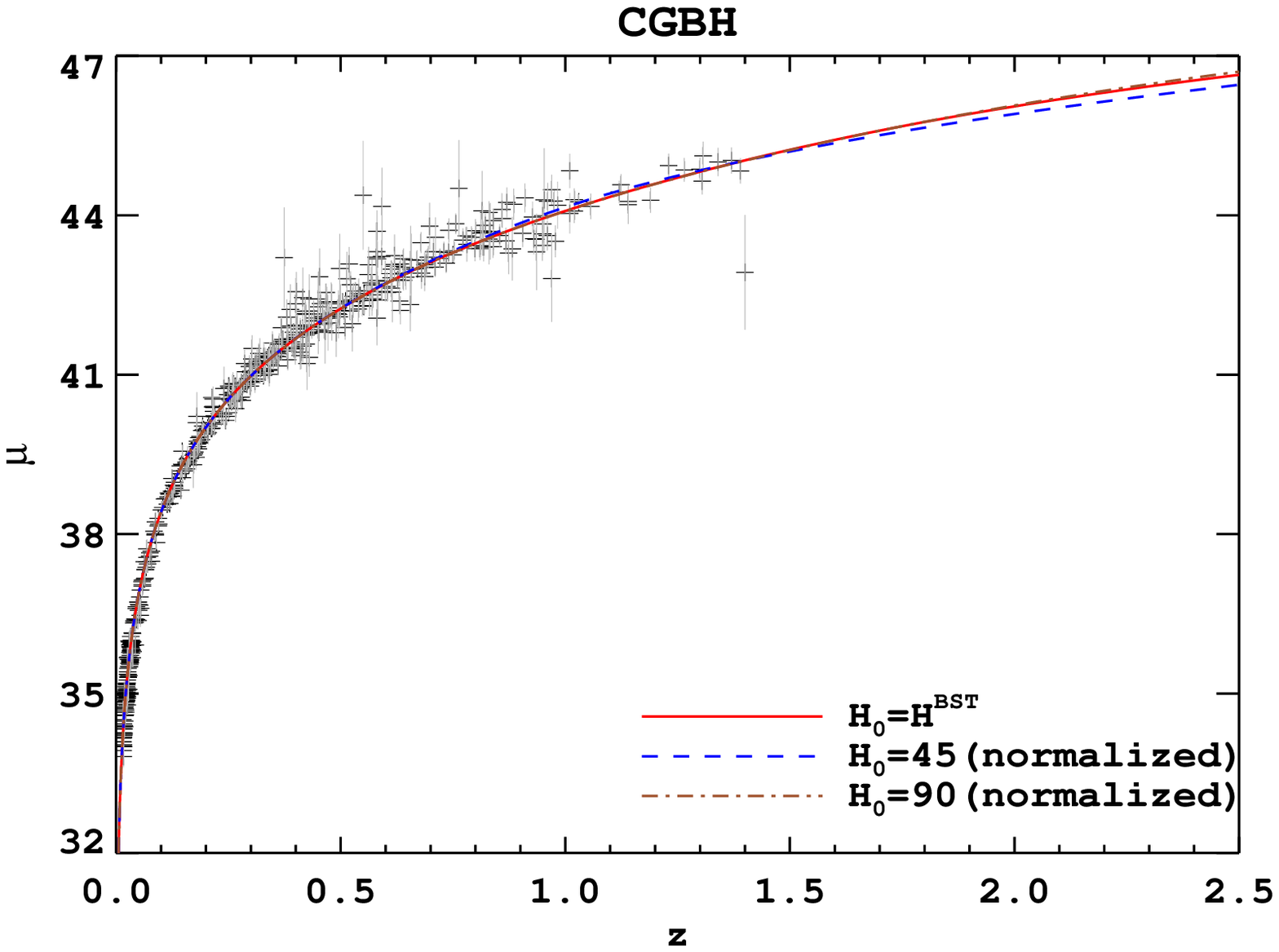}
\includegraphics[width=0.32\textwidth]{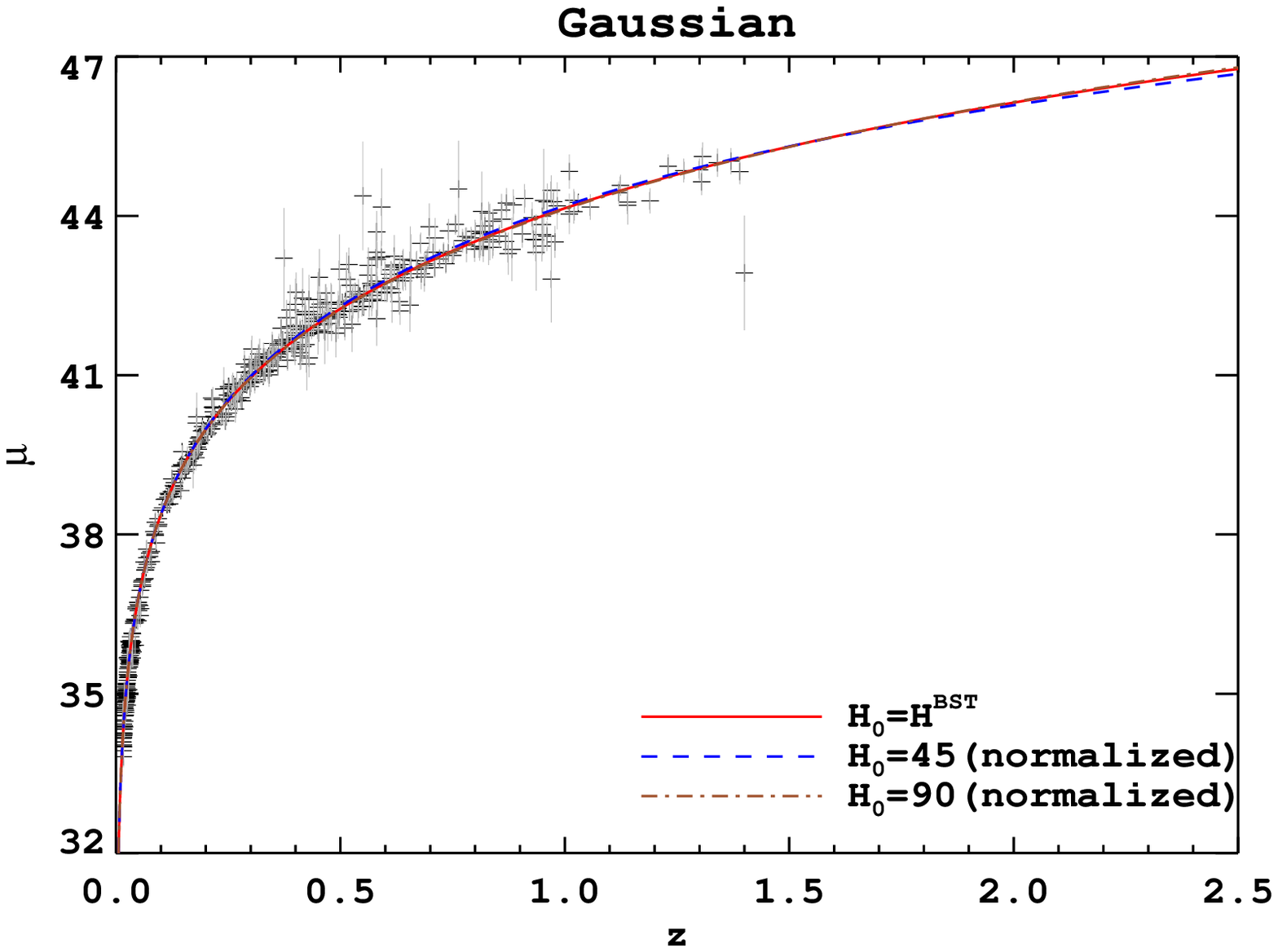}
\includegraphics[width=0.32\textwidth]{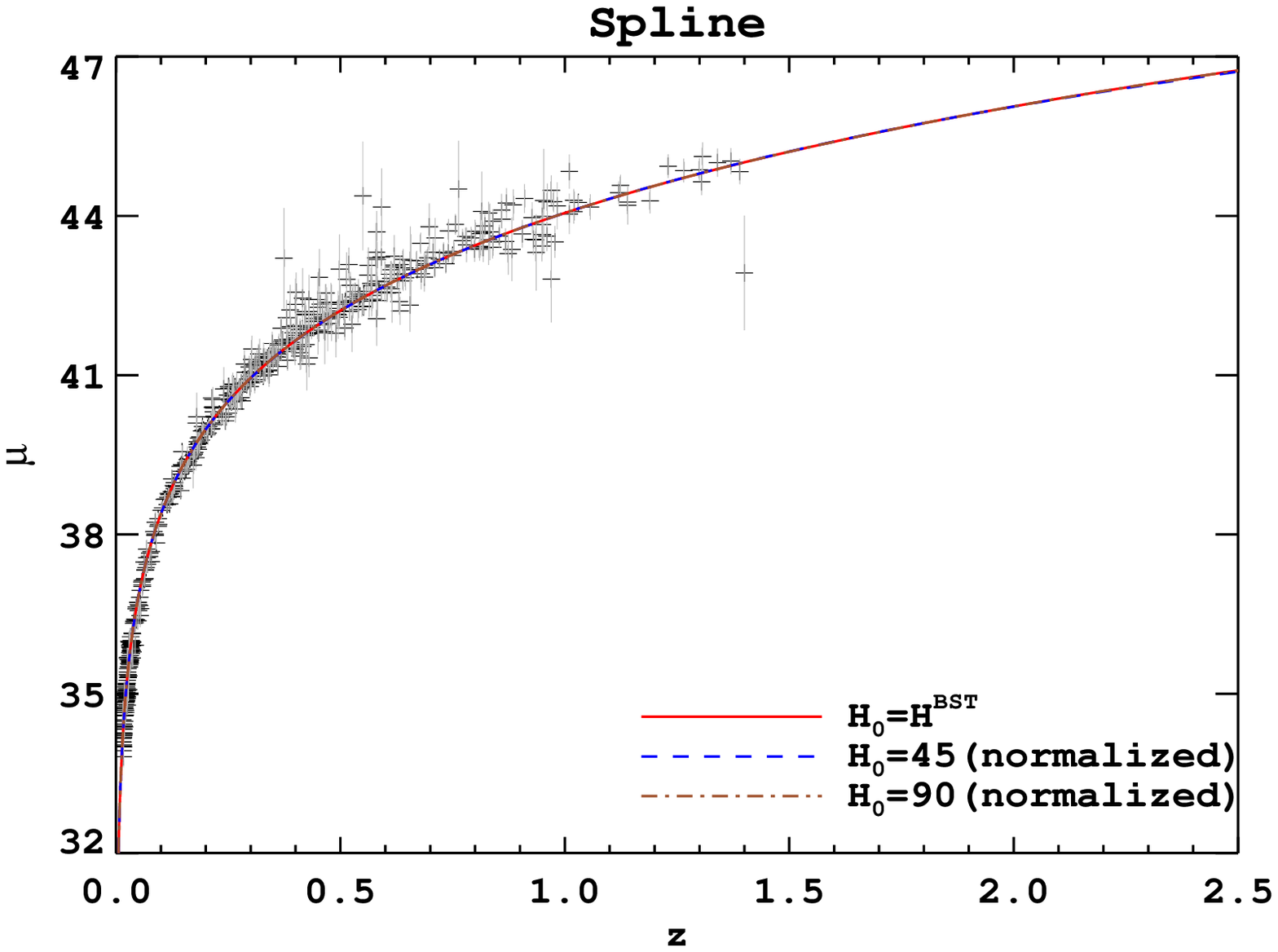}
\end{center}
\caption{Normalized $\mu-z$ relation at different $H_0$. The distance modulus $\mu$ given by $H_0=H$ are normalzed by multiplying the corresponding luminosity distances by $H/H^\text{BST}$, where $H^\text{BST}$ are the best-fit $H_0$ value for each model, other parameters all take the best-fit values determined from the Union2 dataset. The difference between curves of $H=45$, $75$, $90$ are well within the observational uncertainties.
 \label{Fig.hnorm}}
\end{figure}
The observed distance modulus data is given by \citep{Guy2007, Amanullah2010}
\be
\label{Eq.muobs}\mu^{obs}=m_B^*-M+\alpha_x\cdot x_1-\beta\cdot c,
\ee
where $m_B$, $x_1$, $c$ are derived from the fit to the supernovae lightcurves and therefore are known before fitting the cosmology, $M$, $\alpha_x$ and $\beta$ are fitted together with the cosmological parameters to minimize the residuals in the Hubble diagram.

The theoretical distance modulus $\mu$ are derived from the luminosity distance,
\be
\label{Eq.mudl}\mu^{th}=5\text{log}\left(\frac{d_L}{\text{Mpc}}\right)+25.
\ee
In FRW cosmology, $d_L$ is proportional to the reciprocal of the Hubble parameter,
\be
d_L^{\text{FRW}}=(H_0^{\text{FRW}})^{-1}D_L(\theta),
\ee
where $D_L$ is a function of the cosmological components, $\theta$. One can see how $H_0^{FRW}$ degenerates totally in this case with the absolute magnitudes of the supernovae in Eq.\ref{Eq.muobs} by writing explicitly the $\chi^2$ statistic:
\be
\begin{split}
\chi^2(M, \alpha, \beta, H_0^{\text{FRW}}, \theta)
=&\sum\limits_i\frac{\mu^{\text{FRW}}(M, \alpha, \beta, H_0^{\text{FRW}}, \theta, z_i)-\mu^{obs}}{\sigma_i^2}\\
=&\sum\limits_i\frac{m_B^*-5\text{log}_{10}D_L(\theta, z_i)-25+(-M+\alpha\cdot x_1-\beta\cdot c+5\text{log}_{10}H_0^{\text{FRW}})}{\sigma_i^2}.
\end{split}
\ee
This makes it impossible to obtain constraints of $H_0^{\text{FRW}}$ from, say, the Union2 dataset in the FRW framework.

In LTB models, the expansion rate goes into the distance in a more complicated way. Keeping $\Omega(r)$ profile unchanged and varying $H_0$, the pattern of distance modulus-redshifts curve also changes, as is argued and shown in Figure 4 of FLSC. However, we found that in the redshifts range we are concerned, $z\leq2.5$, treating $H_0$ as a normalization factor in all our testing models leads to an error that is too small to be of any importance.

More specifically, we calculate the luminosity distance for each model at their best-fit parameters, then redo the calculation with the value of $H_0$ changed to be $45$ (and then $90$) and other parameters unchanged. The `normalized' luminosity distance are calculated to be
\be
\begin{split}
 d_L^\text{N45}=&d_L^{45}\frac{45}{H_0^\text{BST}}, \\
 d_L^\text{N90}=&d_L^{90}\frac{90}{H_0^\text{BST}},
\end{split}
\ee
respectively, where the superscript `BST' denotes the best-fit value. The normalized distance modulus $\mu^N$are then obtained according to Eq. \ref{Eq.mudl}. The two specific number, $45$ and $90$, are so chosen that the values of $H_0$ scanned in our calculation all lie between them, therefore this choice should serve as an upper bound of the possible error introduced by the approximation. The relative difference in the normalized distance modulus varying $H_0$ from 45 to 90 is
\be
\delta_\mu=\frac{|\mu^\text{N90}-\mu^\text{N45}|}{\mu^\text{N45}}.
\ee
The maximum value of $\delta_\mu$ at $z\leq 2.5$ for the CGBH model is $0.54\%$, Gaussian void $0.26\%$, and Spline void $0.05\%$, respectively, i.e., all less than $1\%$ and can not be differentiated by the current observation (see Figure \ref{Fig.hnorm} for an illustration). Therefore, we conclude that, at least for the models discussed and the redshifts concerned in the current work,  $H_0$ can still be regarded as a normalization factor.

\section{Constraints on the Gaussian and the Spline void model}\label{append.svoid}
\begin{figure*}[!htb]
\begin{center}
\includegraphics[width=0.5\textwidth]{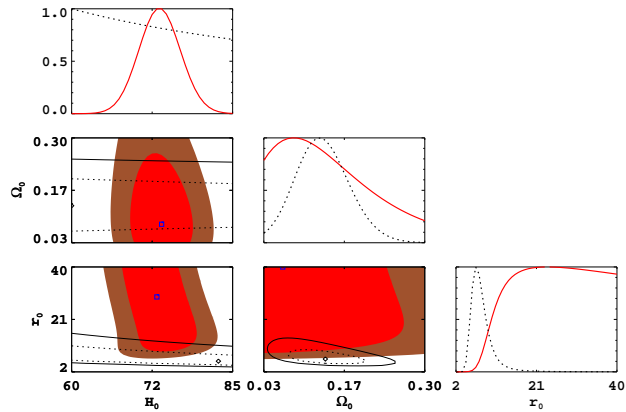}
\end{center}
\caption{Constraints on Gaussian void model parameters from OHD and SN Ia. The red and brown filled regions correspond respectively to 1$\sigma$ and 2$\sigma$ confidence regions obtained from the SJVKS dataset of OHD. Overplotted are the 1$\sigma$ and 2$\sigma$ confidence regions obtained from the Union2 dataset of SN Ia, in solid and dotted lines, respectively.
\label{Fig.gaussvoid}}
\end{figure*}

\begin{figure*}[!htb]
\begin{center}
\includegraphics[width=0.5\textwidth]{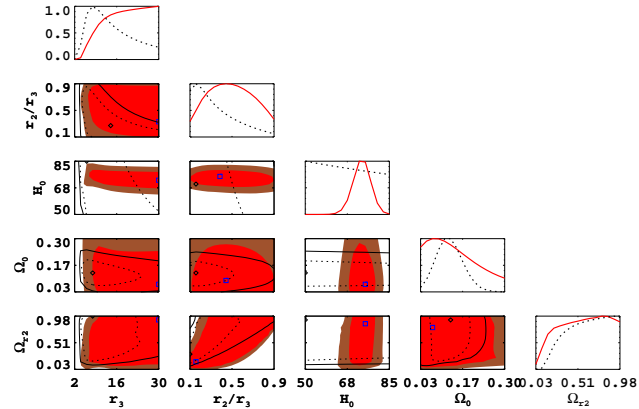}
\end{center}
\caption{Same as Figure \ref{Fig.gaussvoid} but for the Spline void model parameters.
\label{Fig.splinevoid}}
\end{figure*}
The constraints on the Gaussian and the Spline void models are shown in Figure \ref{Fig.gaussvoid} and Figure \ref{Fig.splinevoid}, respectively. The scanning ranges of the parameters are the same as and can be told from the axis in the plots. From the constraints one can tell why the answers to Question-II in Section \ref{sec.forecast} are so different for the Gaussian void and the Spline void models: in the Gaussian void model the $1\sigma$ regions of $r_0-\Omega_0$ from OHD and SN Ia are only slightly overlapped, so a few more data points of OHD will be able to separate them; the confidence regions of $r_2-\Omega_{r2}$ of the Spline void model, on the other hand, contain a much larger mutual part, therefore many more data points and/or much higher precision are needed to separate them.

\end{document}